\documentclass[seceq]{ptptex}
\usepackage{epsfig}

\newcommand{\tr}{{\rm Tr}}
\def\lsim{\raise0.3ex\hbox{$<$\kern-0.75em\raise-1.1ex\hbox{$\sim$}}}
\def\gsim{\raise0.3ex\hbox{$>$\kern-0.75em\raise-1.1ex\hbox{$\sim$}}}

\title{
Heavy quark free energies and the renormalized Polyakov loop in full QCD
}

\author{ 
Olaf \textsc{Kaczmarek}\thanks{Presented by O.~Kaczmarek} with
Shinji \textsc{Ejiri},
Frithjof \textsc{Karsch},
Edwin \textsc{Laermann} and
Felix \textsc{Zantow}
}

\inst{ 
Fakult\"{a}t f\"{u}r Physik, Universit\"{a}t 
Bielefeld, D-33615 Bielefeld, Germany
}

\abst{
We study the renormalized free energy of a heavy quark 
anti-quark pair in the different colour channels 
in full QCD at finite temperature. 
Similarities and differences to the quenched case are discussed and
the temperature dependence as well as 
their short distance behavior are analyzed. 
The asymptotic large distance behavior 
of the free energy is used to 
define the non-perturbatively renormalized Polyakov loop which is 
well behaved in the continuum limit. 
}

\begin{document}

\maketitle

\section{Introduction}
The study of the fundamental forces between quarks is a key to the understanding
of QCD and the occurrence of different phases at high 
temperatures ($T$) and densities ($\mu$). The free energy of a static quark anti-quark pair, 
separated by a distance $r$,
is a good tool to analyze the $r$ and $T$ dependence of the forces,
potentials and entropy contributions of static quarks in the medium
at non-zero $T$ and $\mu$. 
The short and intermediate distance regime of those observables, $rT\lsim 1$, is
relevant for the discussion of in-medium modifications of heavy quark bound
states which are sensitive to thermal modifications of the heavy quark potential.

Here we will discuss first results on the heavy
quark free energies in different color channels in QCD with dynamical quarks.
Results for the quenched theory and a detailed description about the
renormalization procedure can be found in Ref.~1) and Ref.~2).      
While in earlier studies of the heavy quark free energy in full QCD\cite{DeTar:1998qa}
only the color averaged operators were analyzed, we have
a quite detailed description of the different color channels in the
quenched theory.
Below the deconfinement phase transition, the free energies
show the same linearly rising behavior at large distances
governed by the string tension. 
Above $T_c$, the free energies are exponentially screened at large 
distances ($r T \gg 1)$ due to the generation of a chromoelectric
(Debye) mass.
At very small separations, $r T \ll 1$, one gets
into the perturbative regime, where the relevant scale is set by the
distance $r$ and no temperature effects are seen, even at high temperatures
in the deconfined phase. In this region, the singlet
free energy is well described by the zero temperature potential and the
running coupling depends on the dominant scale, i.e. $g=g(r)$.   

In contrast to a linear rising potential in the quenched theory, in QCD with 
dynamical fermions the free energies below $T_c$ show a different behavior at
large separations. Due to the possibility of pair creation the string
between the two {\it test} quarks can {\it break}, leading to a constant free energy
at large separations. 
At very small distances, again the dominant scale is given by the distance $r$ 
and the free energies are expected to be well described by the
zero temperature potential, i.e. zero temperature perturbation theory at
sufficiently small $r$.

Polyakov loop correlation functions are generally used to analyze the 
temperature dependence of confinement forces and the screening in the
high temperature phase of QCD. They are directly related to the {\it change in
  free energy} arising
from the presence of a static quark anti-quark pair in a thermal
medium, $\langle {\rm Tr} W(\vec{x}) {\rm Tr} W^{\dagger}(0) \rangle \sim
\exp(-F_{\bar{q}q}(r=|\vec{x}|,T)/T)$ \cite{Kaczmarek:2002mc,McLerran:pb}~. 

In section 2 we will describe the behavior of the free energy in different color channels
and in the rest of the paper 
we will then
concentrate on results from calculations of the singlet free energy,
$F_1(r=|\vec{x}|,T)/T = -\ln \langle {\rm Tr} W(\vec{x}) W^{\dagger}(0) \rangle$.
The operators used to calculate $F_1$ as well as the octet free energy, $F_8$, are not gauge invariant.
Calculations thus have been performed in Coulomb gauge. It has been 
shown that this approach is equivalent to using a suitably
defined gauge invariant (non-local) operator for the singlet free
energy\cite{Philipsen}~.

\begin{figure}[t]
\centerline{
\epsfig{file=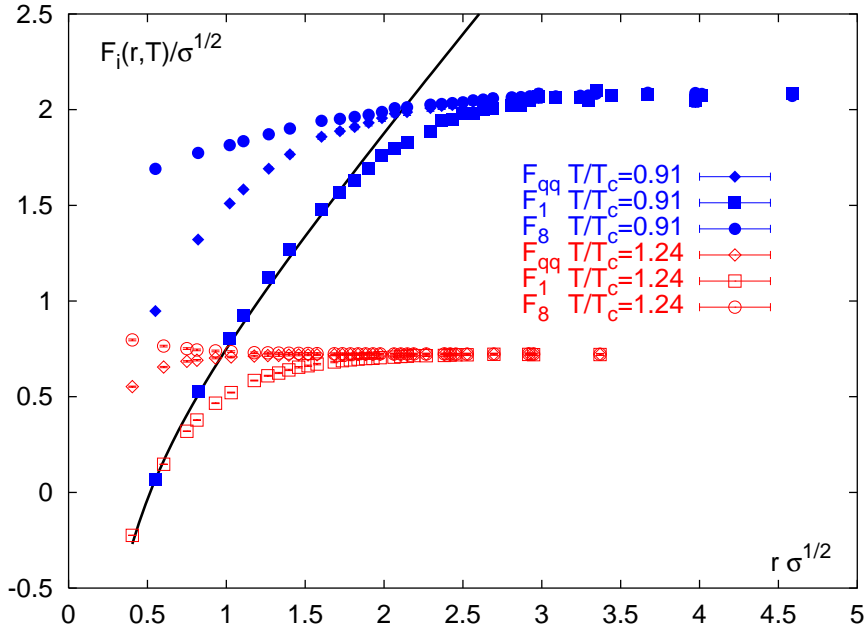,width=12.5cm}
}
\caption{
Heavy quark free energies for 2-flavors of dynamical quarks at a quark mass of $m/T=0.40$ on $16^3\times 4$
lattices. Shown are the different color channels, the singlet ($F_1$), octet
($F_8$) and color averaged ($F_{\bar q q}$) renormalized to the zero-$T$
potential obtained from\cite{Karsch:2000kv}(solid line).
}
\label{saos}
\end{figure}

\section{The renormalized free energy}
We will analyze in the following the properties of heavy quark anti-quark pairs 
in a thermal heat bath of gluons and dynamical quarks.
To analyze the heavy quark free energies we have calculated Polyakov loop 
correlation functions on configurations generated with the p4-action in 
the fermionic sector and a Symanzik improved gauge action in the gauge sector.
On $16^3\times 4$ lattices we have used 2-flavors of dynamical quarks
with a quark mass of $m/T=0.4$.
The zero temperature potential and the temperature scale were obtained from
the data of Ref.~6).

\begin{figure}[t]
\centerline{
\epsfig{file=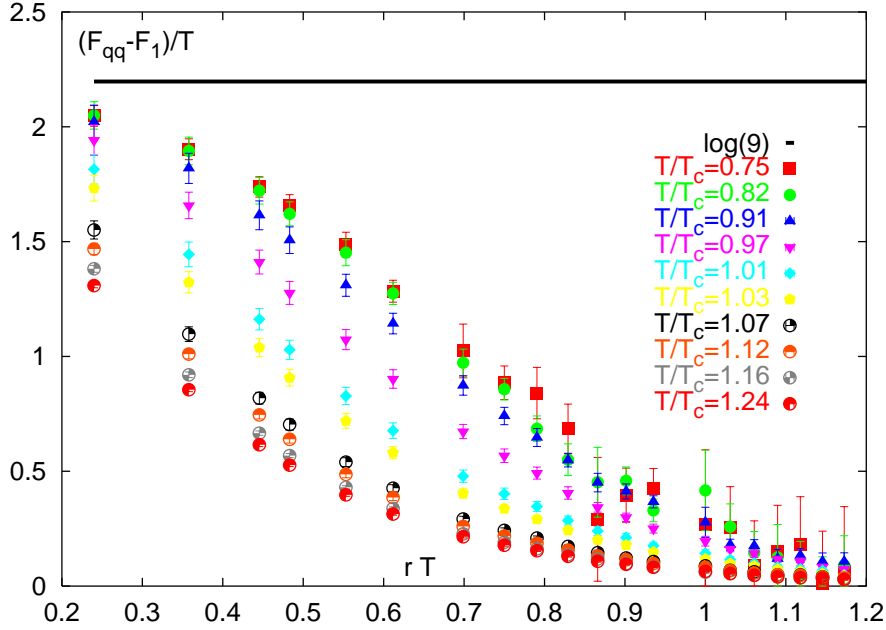,width=12.5cm}
}
\caption{
The difference of the color averaged and singlet free energies for 2-flavors
of dynamical quarks at a quark mass of $m/T=0.40$ on $16^3\times 4$ lattices. The black
line indicates the expected asymptotic short distance limit (solid line).
}
\label{avsingc}
\end{figure}

The static quark sources are described by the Polyakov loop,
\begin{eqnarray}
W(\vec{x}) = \prod_{\tau=1}^{N_\tau} U_0(\vec{x},\tau)
\end{eqnarray} 
with $U_0(\vec{x},\tau) \in SU(3)$ being defined on the link in time direction. The
free energies in the singlet and octet channels are then defined as \cite{Nadkarni:1986cz,McLerran:pb}
\begin{eqnarray}
e^{-F_1(r)/T+C}&=&\frac{1}{3} \tr \langle  W(\vec{x}) W^{\dagger}(0) \rangle 
\label{f1}\\
e^{-F_8(r)/T+C}&=&\frac{1}{8}\langle \tr  W(\vec{x}) \tr W^{\dagger}(0)\rangle- \nonumber \\
&&
                \frac{1}{24} \tr \langle W(\vec{x}) W^{\dagger}(0) \rangle\; ,
\label{f8}
\end{eqnarray}
where $r=|\vec{x}|$ and $C$ is a suitably chosen renormalization constant.
Furthermore, one can consider the color averaged
free energy defined through the correlation function 
\begin{eqnarray}
e^{-F_{\bar q q}(r)/T+C}=\frac{1}{9}\langle \tr W(\vec{x}) \tr W^{\dagger}(0) \rangle \; .
\label{f}
\end{eqnarray}
This can be written as thermal average over free energies in
singlet and octet channels
\begin{eqnarray}
e^{-F_{\bar q q}(r)/T}=\frac{1}{9} e^{-F_1(r)/T}+\frac{8}{9} e^{-F_8(r)/T}.
\label{f18}
\end{eqnarray}
At distances much shorter than the inverse temperature ($r T \ll 1$) the
dominant scale is set by $r$ and the running coupling will be controlled by
this scale and become small for ($r \ll 1/\Lambda_{QCD}$). In this limit
the singlet and octet free energies are dominated by one-gluon exchange and
become calculable within ordinary zero temperature perturbation theory, i.e.
are given by the singlet and octet heavy quark potential.
We have used this to fix the constant $C$ in (\ref{f1}),(\ref{f8}) and (\ref{f18}) by matching the
singlet free energy to the zero temperature heavy quark potential at short
distances. 

\begin{figure}[t]
\centerline{
\epsfig{file=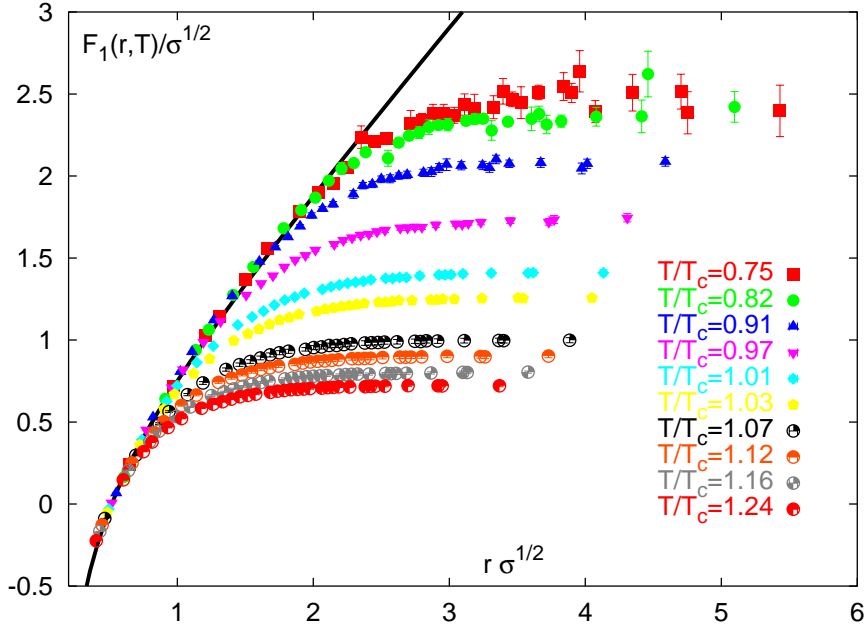,width=12.5cm}
}
\caption{
Heavy quark free energies in the singlet channel 
for 2-flavors of dynamical quarks at a quark mass of $m/T=0.40$ on $16^3\times 4$
lattices renormalized to the zero-$T$
potential obtained from\cite{Karsch:2000kv}(solid line).
}
\label{plcs}
\end{figure}

In fig.~\ref{saos} the renormalized free energies in the different color
channels for two temperatures are plotted. At small distances $F_1$
coincides with the $T$=0-potential. 
For the temperature of $0.91~T_c$ we
see no thermal effect up to a distance of $r\sqrt{\sigma}\approx 1.5$ 
where string breaking sets in and leads to a constant value at larger
separations.
For $T$=$1.24~T_c$ the thermal effect sets in at
$r\sqrt{\sigma}\approx 0.7$. The singlet free energy, $F_1$, shows the usual
screened Coulomb
like behavior approaching a temperature dependent constant value at large distances.
In all color channels the free energies reach the same constant ({\it
  cluster}) value at large separations above as well as below $T_c$.

While the singlet potential is attractive, the octet potential is repulsive at
short distances. From eq.~(\ref{f18}) it follows that in this limit the color
averaged free energy will be dominated by the singlet contribution. We may then
deduce from (\ref{f18}) also the asymptotic short distance behavior of $F_{\bar
  q q}$ and $F_1$,
\begin{eqnarray}
\lim_{r\rightarrow 0} (F_{\bar q q}(r,T) - F_1(r,T)) = T \ln 9 \ \ \
\mathrm{for \ all \ }T.
\label{ln9ref}
\end{eqnarray}
In fig.~\ref{avsingc} this asymptotic behavior at small distances is
reached, although for the higher temperatures
the distances analyzed here are not small enough to get into the regime where
(\ref{ln9ref}) is fulfilled. At large distances the difference between the
singlet and color averaged free energies vanishes and therefore  
$F_1$, $F_{\bar q q}$ and consequently also $F_8$ reach 
the same {\it cluster} value at large separations. 

\section{Short vs. long distances}

In the following we will concentrate on the singlet free energies
and analyze their short and long distance behavior. In fig.~\ref{plcs} we show $F_1(r,T)$
in the temperature range of $0.75 < T/T_c < 1.25$. We see again the 
$T$-independence at sufficiently small distances. The thermal effects set
in at relatively large distances of $r\sqrt{\sigma}\approx 2$
for the lowest
temperatures and move towards smaller distances with increasing $T$.

At large distances $F_1$ is {\it screened} and reaches constant values for all
temperatures, which can be explained by string breaking below $T_c$ and 
screening due to the generation of screening masses
above the transition temperature.

\begin{figure}[t]
\centerline{
\hspace*{-0.1cm}\epsfig{file=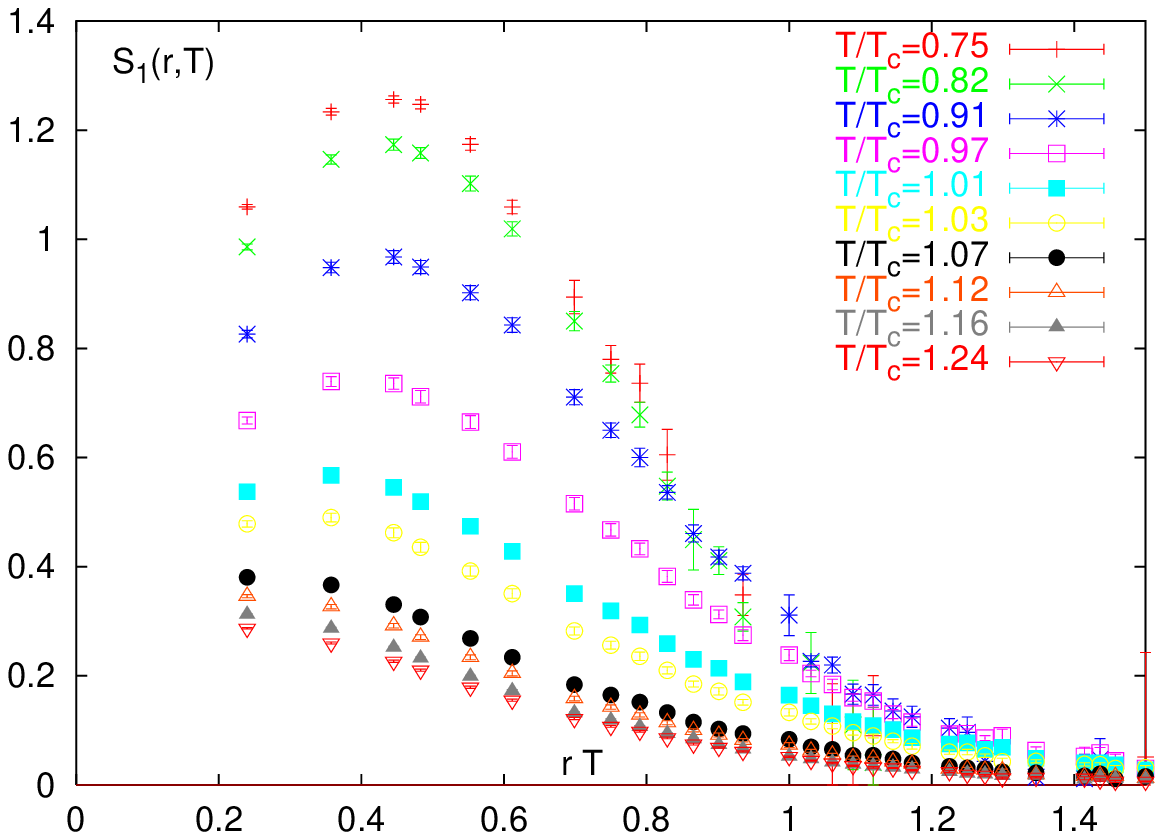,width=7.5cm}\hspace*{-0.2cm}\epsfig{file=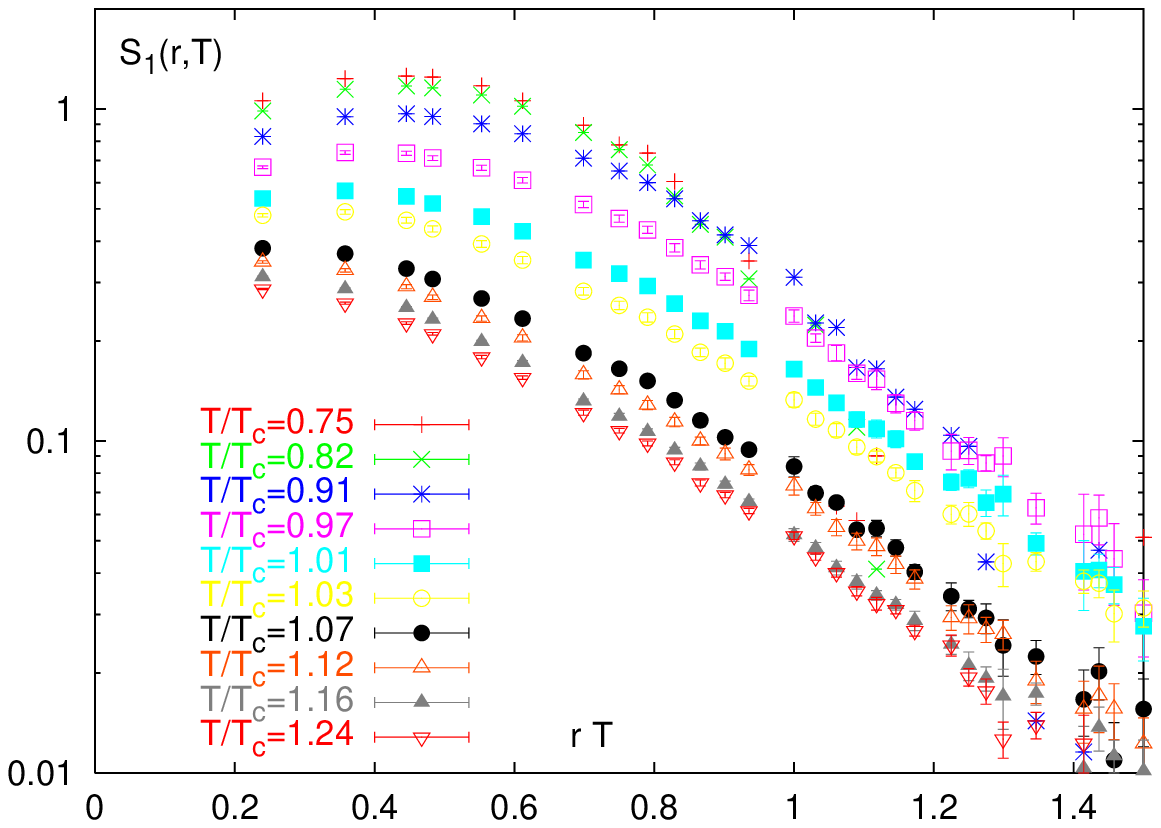,width=7.5cm}
}
\caption{
The color singlet screening function on linear (left) and logarithmic (right) scales.
}
\label{screenf}
\end{figure}

To analyze at what distances these screening effects set in, we introduce the
color singlet screening function as
\begin{eqnarray}
S_1(r,T) = -\frac{3}{4} r (F_1(r,T)-F_1(\infty,T)).
\label{log9f}
\end{eqnarray}
At short distances this quantity should approach the running coupling constant,
$\alpha=g^2/4\pi$, and thus is expected to drop logarithmically, while at large
distances it carries information about screening and is expected to drop
exponentially. Consequently we expect that $S_1(r,T)$ will exhibit a maximum at
some intermediate distance which we can identify as the point separating the
short distance physics from the large distance regime.

Fig.~\ref{screenf} shows the data on $S_1(r,T)$ on both linear (left) and
logarithmic (right) scales. At small temperatures a clear maximum is
visible at distances of $r T\simeq 0.45$. At higher temperatures
indications for a tendency
to develop a maximum are visible, but the distances analyzed here are not small
enough to verify this. 
We find that the maximum, i.e. the point separating the short from the 
large distance regime, occurs at $rT\simeq 0.45$ at $0.75 T_c$ and slowly
shifts to smaller values at high temperatures, e.g. at $T_c$ it is at
$rT\simeq 0.4$.
Beyond this length scale $S_1(r,T)$ drops rapidly and thus exhibits
screening. The distances analyzed here are too small to really see the simple
exponential form, as distances $rT\geq 1$ are needed to separate the long
distance regime from the short distance one.

The strong thermal effects seen in fig.~\ref{screenf} even at small
distances are to a large extend caused by our normalization in (\ref{log9f}),
which forces $S_1(r,T)$ to approach zero at large separations, but introduces
an artificial temperature dependence at short distances. 

Looking at the temperature dependence of the singlet free energies in fig.~\ref{plcs},
$F_1(r,T)$ decreases with increasing temperature at fixed $r$, indicating that there
is a positive entropy, $S=-\frac{\partial F_1}{\partial T}$,  contribution at
large distances, while it is close to zero,
due to the (asymptotic) $T$-independence of $F_1(r,T)$, at small $r$.

\begin{figure}[t]
\centerline{
\hspace*{-0.5cm}\epsfig{file=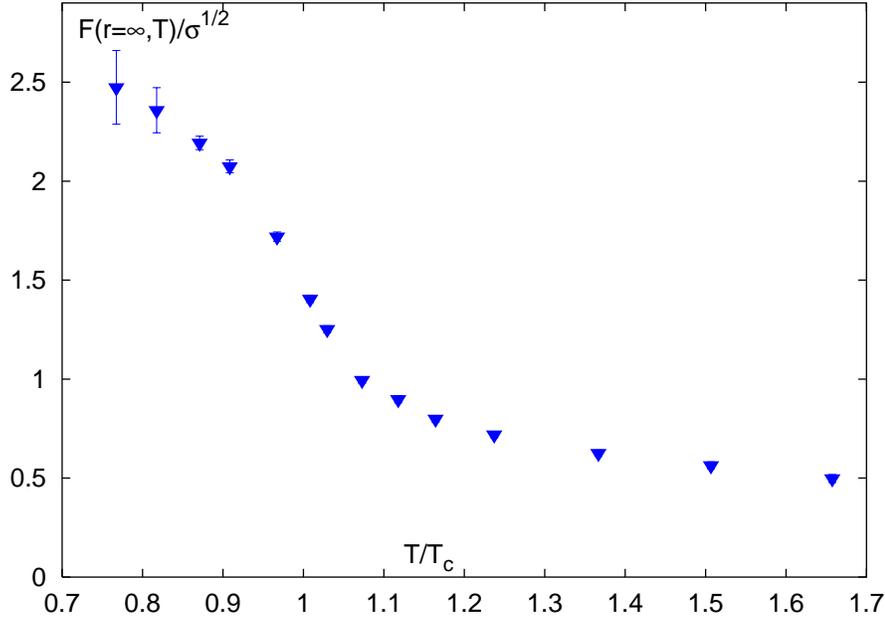,width=12.5cm}
}
\caption{
The asymptotic large distance values of the free energies.
}
\label{finf}
\end{figure}

\begin{figure}[t]
\centerline{
\hspace*{-0.5cm}\epsfig{file=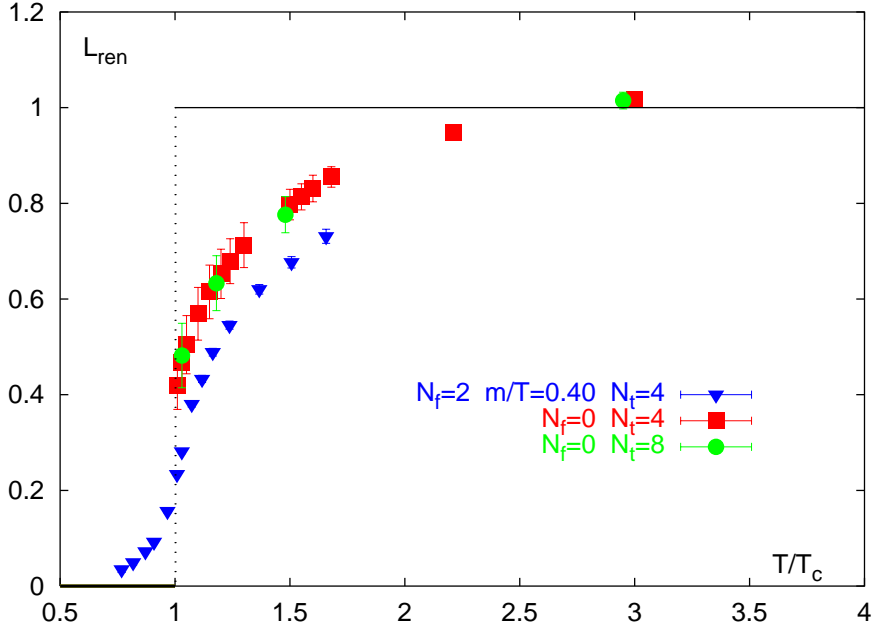,width=12.5cm}
}
\caption{
The renormalized Polyakov loop in full QCD compared to the
quenched results\cite{Kaczmarek:2002mc}~.
}
\label{renpol}
\end{figure}

\section{The renormalized Polyakov loop}
The Polyakov loop, calculated on the lattice, is ultra-violet divergent
and needs to be renormalized to become a meaningful observable in
the continuum limit. We will do so by renormalizing the free energies
at short distances. Assuming that no additional divergences arise
from thermal effects and that at short distances the heavy quark
free energies will not be sensitive to medium effects, renormalization is achieved 
through a matching of free energies to the zero temperature heavy quark 
potential. Using the large distance behavior of the renormalized
free energies we can then define the renormalized Polyakov loop which is well
behaved also in the continuum limit.  

Using the renormalized free energies
from fig.~\ref{plcs}, i.e. the asymptotic values in fig.~\ref{finf}, we can define the renormalized Polyakov loop
\cite{Kaczmarek:2002mc}~,

\begin{eqnarray}
L_{\mathrm ren} = \exp\left( - \frac{F_1(r=\infty,T)}{2 T}\right).
\label{renpo}
\end{eqnarray}

In fig.~\ref{renpol} we show the results for $L_{\mathrm ren}$ in full QCD
compared
to the quenched results obtained from Ref.~1).
In quenched QCD it is zero below $T_c$ by construction, as the free energy
goes to infinity in the limit of infinite distance. From the results of
different values of $N_\tau$, it is apparent that $L_{\mathrm ren}$ does not depend on
$N_\tau$ and therefore is well behaved in the continuum limit.

The renormalized Polyakov loop in full QCD
is no longer zero below $T_c$. Due to 
string breaking the free energies reach a constant value at large separations 
leading to a non-zero value of $L_{\mathrm ren}$. The renormalized
Polyakov loop is no longer an order parameter for finite quarks mass, but 
still indicates a clear signal for a phase change at $T_c$. It is small
below $T_c$ and shows a strong increase close to the critical temperature. 
In the temperature range we have analyzed, $L_{\mathrm ren}$ is smaller
in full QCD compared to the quenched case.
In Ref.~6)
no major quark mass effects were visible in the color
averaged free energies below 
a quark mass of $m/T=0.4$. To verify this for the singlet
and octet channel, a more detailed analysis of the mass and also 
flavor dependence is needed.

\section{Conclusions}
We have discussed the renormalized free energies in the different color
channels for 
QCD with dynamical quarks. 
The results in 2-flavor QCD show that the concepts, developed in quenched QCD,
can be extended to full QCD. A more precise analysis of the temperature and
mass (and flavor) dependence, as well as a closer look at the short distance regime, is needed.
For a detailed analysis of screening phenomena at high temperatures, larger
lattices are needed, as separations of $rT\geq 1$ are required to obtain
screening masses from the exponential fall off. We have shown that there is
a quite different behavior for short and large distances and that those
regimes are separated around $rT\simeq 0.4$ near $T_c$.
From the analysis of the screening function we find
indications of the running of the coupling at small distances, but
smaller lattice cut-offs are needed to analyze this behavior in more detail.

The large distance behavior of the free energy was used to calculate the
renormalized Polyakov loop. We observe visible differences to the quenched
results. 
$L_{ren}$ is no longer zero below $T_c$, but 
still indicates different behavior in both phases and a strong increase  
close to $T_c$.
It is smaller compared to the quenched case in the
temperatures region we have studied.

The extension of the analysis described here to non-zero density using a
Taylor expansion \cite{Allton:2003vx} in $\mu$ is straightforward and in progress.

\section*{Acknowledgments}

We would like to thank the Bielefeld-Swansea collaboration for providing their
configurations.
This work is supported by 
BMBF under grant No.06BI102 and DFG under grant FOR 339/2-1.

\end{document}